\documentclass[prb,twocolumn,floatfix,superscriptaddress,tightenlines]{revtex4-1}
\usepackage{graphicx,epsfig}
\usepackage{hyperref}
\begin{document}
\title{Second and higher harmonics generation with memristive systems}
\author{Guy Z. Cohen}
%\email{gcohen@physics.ucsd.edu}
\affiliation{Department of Physics, University of California, San
Diego, La Jolla, California 92093-0319}
\author{Yuriy V. Pershin}
%\email{pershin@physics.sc.edu}
\affiliation{Department of Physics and Astronomy and USC
Nanocenter, University of South Carolina, Columbia, SC, 29208}

\author{Massimiliano Di Ventra}
%\email{diventra@physics.ucsd.edu}
\affiliation{Department of Physics, University of California, San
Diego, La Jolla, California 92093-0319}
\begin{abstract}
We show that memristive systems can be used very efficiently to
generate passively both double and higher frequency harmonics. A
technique for maximizing the power conversion efficiency into any given
harmonic is developed and applied to a single memristive system
and memristive bridge circuits. We find much
higher rates of power conversion compared to the standard diode bridge, with the memristive bridge
more efficient for second and higher harmonics generation compared to the single memristive system.
The memristive bridge circuit optimized for second harmonic generation behaves as a two-quarter-wave rectifier.
\end{abstract}
\maketitle
\thispagestyle{plain} In nonlinear optics, the phenomenon of
second-harmonic generation (SHG), demonstrated in
1961~\cite{Franken61a}, refers to the possibility of creating an
outgoing light beam with double the frequency of the impinging
beam~\cite{Hecht01a}. This is usually achieved with the help of
nonlinear crystals. The smallness of the nonlinearity, however,
limits the efficiency of direct SHG. The use of lenses and mirrors
improves the efficiency up to 85\% by making the light pass
repeatedly through a nonlinear medium~\cite{Ou92a}. A common
application of SHG in optics is the creation of laser sources with
modified frequencies.

In the field of electronics, SHG is termed as frequency doubling.
Electronic frequency doublers can be divided into active
ones, requiring an external power source, and passive ones,
operating exclusively on the power of the input signal (see, e.g.,
Ref. \onlinecite{pershin09a}). A straightforward realization of a
passive frequency doubler is based on a diode
bridge~\cite{Alexander08a}, which produces a full-wave rectified
output (mathematically, the absolute value of a sinusoidal input).
Using Fourier analysis, it can be shown that the diode bridge
transforms 4.50\% of the input power into the second harmonic, and
18.9\% of the input power into all higher
harmonics~\cite{Alexander08a}. In addition, by symmetry its output
signal contains only even harmonics.

Memristors~\cite{Chua71a}, which belong to the more general class
of memristive systems~\cite{chua76a}, are resistive circuit
elements with memory. A recent understanding~\cite{Strukov08a} of
the memristive nature of resistance switching in memory cells has
attracted a lot of attention to the field of memory
elements~\cite{pershin11a}, which comprises also memcapacitive and
meminductive systems~\cite{diventra09a}, namely, capacitors and
inductors with memory.

In this paper we demonstrate that a passive circuit with a {\it single} memristive device generates
 second and higher harmonics signals on a load resistor with significantly higher efficiency than that of the diode bridge. Moreover, the efficiency
of harmonics generation can be further improved employing a
\emph{memristive bridge} introduced in this paper. We note that
SHG in large memristor networks was anticipated in Ref.
\onlinecite{Oskoee11} but no detailed analysis was given.

\begin{figure}[b]
\vspace{-5pt}
%\begin{center}
\includegraphics[width=8.6cm]{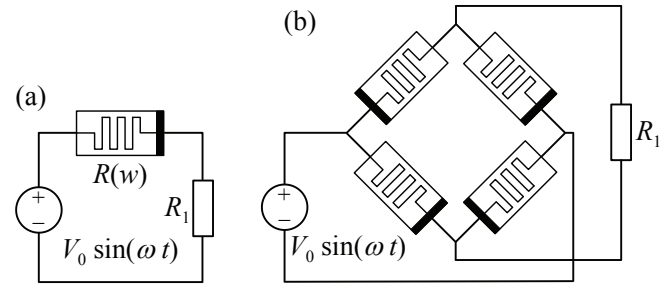}
%\end{center}
\caption{\label{fig01} (a) Circuit consisting of a time-dependent
voltage source  connected to a memristive system with memristance
$R(w)$ and a load resistor $R_1$. (b) Memristive bridge circuit.}
\end{figure}

We start by showing that memristive systems indeed generate second and higher harmonics using a simple model of resistance switching introduced in Ref.
\onlinecite{Strukov08a} to describe the response of a thin TiO$_2$ film
 of a width $D$ sandwiched between two Pt electrodes. The phenomenon of
 resistance switching in this system is understood~\cite{yang08a}
as a field-induced migration of oxygen vacancies changing the width $w$ of
 a doped low-resistance region. Electrical current in the positive direction
(towards the thick line in the circuit symbol of the memristor in
Fig. \ref{fig01}(a)) increases $w$, decreasing the resistance. If
the maximum and minimum resistance values, obtained at $w=0$ and
$w=D$, are denoted by $R_{\mathrm{off}}$ and $R_{\mathrm{on}}$,
respectively, and the charge flow through the device needed to
completely switch it from one limiting state into another is
$q_0$, then the device memristance (memory resistance) can be
written as \cite{Strukov08a}
\begin{equation}
R(w)=R_{\mathrm{off}}+\left (R_{\mathrm{on}}-R_{\mathrm{off}}
\right )\frac{w}{D}, \label{eq01}
\end{equation}
where $w/D=q/q_0$, and $q$ is the cumulative charge flowing through the device.

We first consider a sine voltage source of period $T$ connected in
series to a memristive device $R(w)$ and a load resistor $R_1$
(Fig. \ref{fig01}(a)). Kirchhoff's voltage law for this circuit is
\begin{equation}
V_0\sin\omega t-\dot{q}[R(w)+R_1]=0, \label{eq02}
\end{equation}
where $R(w)$ is given by Eq. (\ref{eq01}) and $\omega\equiv
2\pi/T$. Solving Eq. (\ref{eq02}) with  $q(t=0)=0$, we find the current
\begin{eqnarray}
I(t)=\frac{V_0 \sin(\omega
t)/(R_{\mathrm{off}}+R_1)}{\sqrt{1-\frac{4R_2(R_{\mathrm{off}}-R_{\mathrm{on}})}{(R_{\mathrm{off}}+R_1)^2}\sin^2(\frac{\omega
t}{2})}}, \label{eq03}
\end{eqnarray}
where $I(t)\equiv \dot{q}$, and $R_2\equiv V_0/(q_0\omega )$ is a
constant with dimensions of resistance. First, we see that the
current $I(t)$ in Eq. (\ref{eq03}) has the same periodicity as the
source and, therefore, can be written as a Fourier series. Second,
the numerator in Eq. (\ref{eq03}) is a static resistivity current
expression, while the denominator contains corrections due to
memory. These corrections give rise to all higher harmonics. In
many memristive devices $R_{\mathrm{on}}\ll R_{\mathrm{off}}$, and
 it is reasonable to assume $R_1<R_{\mathrm{off}}$.
Under such assumptions, it follows from Eq. (\ref{eq03}) that the
condition for considerable generation of higher harmonics is
having $R_2/R_{\mathrm{off}}$ of order one (while keeping the
denominator real). For a given memristor, the ratio
$R_2/R_{\mathrm{off}}$ can be increased either by increasing the
amplitude of the source, $V_0$, or by decreasing its frequency,
$\omega/2\pi$.

The memristor model given by Eq. (\ref{eq01}) combined with
$w/D=q/q_0$ is convenient for analytical calculations. However, it
does not limit $R(w)$ between $R_{\mathrm{on}}$ and
$R_{\mathrm{off}}$. In order to obtain
quantitative results, we then suggest a more realistic model
 consisting of Eq. (\ref{eq01}) and
\begin{equation}
\dot{w}=\frac{D}{q_0}\left
[\theta(1-w/D)\theta(\dot{q})+\theta(w/D)\theta(-\dot{q})\right
]\dot{q}, \label{eq01-1}
\end{equation}
where the $\theta$ functions in Eq. (\ref{eq01-1}) constrain $w$
to satisfy $0\leq w\leq D$, in agreement with experimental
data~\cite{Baikalov03a,Tsui04a}.

It follows from Eqs. (\ref{eq01-1}) and (\ref{eq01}) that
$\dot{w}$ has the periodicity of $\dot{q}$, and that the current
$I(t)$ has the same periodicity as that of the voltage source.
However, $I(t)$ cannot be found in a closed analytical form now,
and we solve the problem numerically. If we denote the average
power dissipated on the load resistor in the $i$-th harmonic as
$P_i$ and the average power of the source as
$P_{\mathrm{source}}$, three important ratios come to mind:
$P_2/P_{\mathrm{source}}$, $P_3/P_{\mathrm{source}}$ and
$(\sum_{k=2}^{\infty} P_{k})/P_{\mathrm{source}}$, which are the
second, third and higher harmonics conversion efficiencies,
respectively. Identifying the system parameters maximizing
 these ratios will enable us to generate, upon
application of a suitable band pass filter at the output, the
desired harmonics passively and with minimal losses.

Considering all possible parameters, we can write the general
functional dependence of the above ratios as
\begin{eqnarray}
\frac{P_k}{P_{\mathrm{source}}}&=&f_k\left(\omega,V_0,q_0,w(t=0),R_{\mathrm{off}},R_{\mathrm{on}},R_1\right),\label{eq07} \\
\frac{\sum_{k=2}^{\infty}
P_{k}}{P_{\mathrm{source}}}&=&g\left(\omega,V_0,q_0,w(t=0),R_{\mathrm{off}},R_{\mathrm{on}},R_1\right).
\label{eq08}
\end{eqnarray}
As noted above, these ratios cannot be written in closed
analytical form. Moreover, the numerical maximization is difficult
due to the large number of parameters. It is shown below that it
is reasonable to select $w(t=0)=0$. The number of parameters can
be further reduced utilizing the Buckingham $\pi$ theorem of
dimensional analysis~\cite{Munson10a}. Its application results in
functions of dimensionless variables for the above ratios:
\begin{eqnarray}
\frac{P_k}{P_{\mathrm{source}}}&=&\widetilde{f}_k\left(\frac{R_1}{R_{\mathrm{off}}},\frac{R_2}{R_{\mathrm{off}}},\frac{R_{\mathrm{off}}}{R_{\mathrm{on}}}\right),\label{eq09} \\
\frac{\sum_{k=2}^{\infty}
P_{k}}{P_{\mathrm{source}}}&=&\widetilde{g}\left(\frac{R_1}{R_{\mathrm{off}}},\frac{R_2}{R_{\mathrm{off}}},\frac{R_{\mathrm{off}}}{R_{\mathrm{on}}}\right).
\label{eq10}
\end{eqnarray}
\begin{figure}[t]
 \begin{center}
 \includegraphics[width=8.6cm]{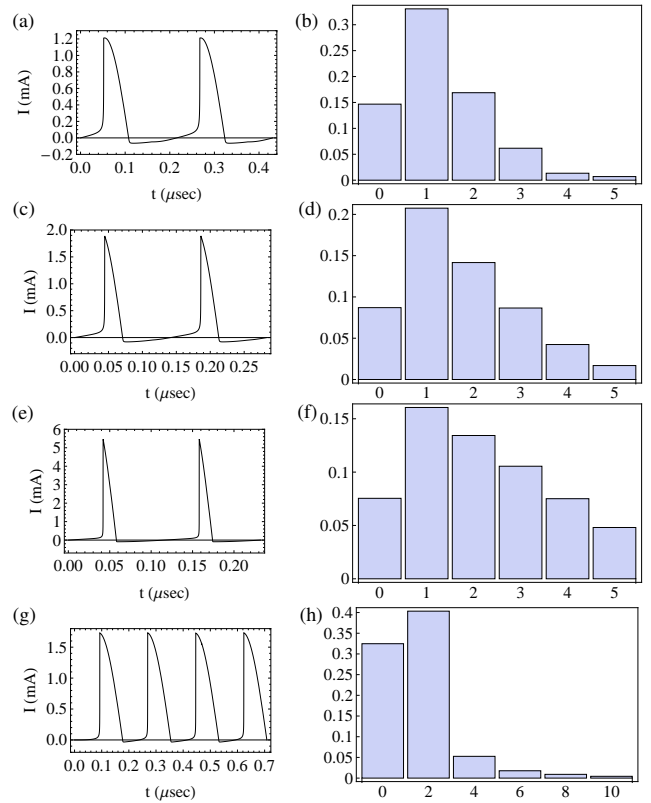}\vspace{-5pt}
 \caption{\label{fig02} (color online) Current $I(t)$
and corresponding power ratio distribution
($P_i/P_{\mathrm{source}}$) after maximization of the power
conversion into the second (a,b), third (c,d), and all higher
harmonics (e,f) in the single memristive device circuit and into
the second harmonic in the memristive bridge circuit (g,h). $I(t)$
has been obtained using $V_0=1$V, $q_0=3\cdot 10^{-12}$C and
$R_{\mathrm{off}}=20$k$\Omega$.}
 \end{center}
   \vspace{-20pt}
\end{figure}

The left-hand sides of Eqs. (\ref{eq09}) and (\ref{eq10}) are
maximized numerically considering  the following realistic ranges
of dimensionless parameters~\cite{Strukov08a}: $10^{-5}\leq
R_1/R_{\mathrm{off}}\leq 10$, $10^{-4}\leq
R_2/R_{\mathrm{off}}\leq 10$ and $10\leq
R_{\mathrm{off}}/R_{\mathrm{on}}\leq 10^3$. We find that the
maxima of $P_2/P_{\mathrm{source}}$ and $P_3/P_{\mathrm{source}}$
depend weakly on $R_{\mathrm{off}}/R_{\mathrm{on}}$ and lie in the
range $150<R_{\mathrm{off}}/R_{\mathrm{on}}<1000$. Consequently,
we set $R_{\mathrm{off}}/R_{\mathrm{on}}=200$. For
$\sum_{k=2}^{\infty} P_{k}/P_{\mathrm{source}}$ it is found that
$R_{off}/R_{on}=1000$ gives the best value.

Table \ref{table1} presents conversion efficiencies and lists the
optimal parameter values. Clearly, in all cases the single
memristive device circuit provides significantly higher conversion
rates than the diode bridge. Figs. \ref{fig02}(a)-(f) show the
current through the load resistor and corresponding harmonics
power spectrum. Since the memristive device is a {\it dynamically
adaptive system},
 $I(t)$ is very different from the sine shape of the source.

\begin{table}[tb]
\begin{tabular}{c | c c c | c c c}
Optimized  & $\frac{R_1}{R_{\mathrm{off}}}$ &
$\frac{R_2}{R_{\mathrm{off}}}$ &
 $\frac{R_{\mathrm{off}}}{R_{\mathrm{on}}}$ & Rate, & Rate, & Rate,\\
 quantity &&&& 4 diodes & 1 memr. & 4 memr.\\
\hline
$P_2/P_{\mathrm{source}}$ & 0.0362 & 1.71 & 200 & 4.50\% & 16.9\% & 40.3\%\\
$P_3/P_{\mathrm{source}}$ & 0.0194 & 1.13 & 200 & 0\% & 8.66\% & 0\%\\
$\frac{\sum_{k=2}^\infty P_k}{P_{\mathrm{source}}}$ & 0.00598 & 0.924 & 1000 & 18.9\% & 48.6\% & 56.4\%\\
\end{tabular}
\caption{Optimized parameter values for single memristive device
circuit, power conversion rates for diode bridge (Rate, 4 diodes),
optimized power conversion rates for single memristive device
(Rate, 1 memr.) and memristive bridge (Rate, 4 memr.) circuits.}
\label{table1}
  \vspace{-15pt}
\end{table}

It is also important to check how the initial value of $w(t=0)$
affects the higher harmonics generation. We have performed
extensive numerical simulations and found that the power conversion rates (Eqs. (\ref{eq07}) and (\ref{eq08}))
always increase with a decrease of $w$. This observation
justifies our choice of $w(t=0)=0$ corresponding to $R(w(t=0))=R_{off}$.
Moreover, we note that once a certain $R(w(t=0))$ is selected, the value of $R_{off}$
becomes not important if $R(w(t))>R_{on}$.
In order to prove this statement, we notice that the load current
has the same period as the source, which
implies that the memristance also has such periodicity. The
memristance increases/decreases with negative/positive current,
which changes sign only at $t=nT$ and $(n+1/2)T$, where $n$ is an
integer. Since the memristance decreases immediately after $t=0$,
we conclude the memristance has maxima at $t=nT$, minima at
$(n+1/2)T$ and no other extrema. Therefore, $R_{on} < R(w)\leq
R(w=0)$ during the entire period, implying that having $w_1\equiv
w(t=0)>0$ and off resistance $R_{off}$ is equivalent to having
$w(t=0)=0$ and off resistance $R(w_1)$. The decrease in the
memristance range reduces the circuit nonlinearity, translating
into lower efficiency of higher harmonics generation.

The efficiency of the second and higher harmonics generation can
be increased even further by utilizing a memristive bridge having
the geometry of the diode bridge (Fig. \ref{fig01}(b)). The
memristive bridge consists of four identical memristive devices
described by Eqs. (\ref{eq01}) and (\ref{eq01-1}) rectifying the
input signal via the delayed switching effect~\cite{Wang10b}. This
rectification reduces the load current periodicity to $T/2$,
leaving only the even harmonics. Consequently, the second and
higher harmonic generation efficiencies are improved relative to
those of the single device circuit, while third harmonic
generation is excluded.

In complete analogy with the single device circuit, dimensional
analysis for the memristive bridge leads to Eqs. (\ref{eq09}) and
(\ref{eq10}) for the efficiency of harmonics generation.
Dimensionless parameters maximizing $P_2/P_{\mathrm{source}}$ are
$R_2/R_{\mathrm{off}}=2.82$, $R_1/R_{\mathrm{off}}=0.0267$ and
$R_{off}/R_{on}=1000$ giving $P_2/P_{\mathrm{source}}=40.3\%$, a
substantial improvement over the single device circuit, with same
initial conditions. $I(t)$ and the power distribution of harmonics
for the memristive bridge are given in Figs. \ref{fig02}(g,h). It
is found that with an increase in $R_{off}/R_{on}$, the optimized
current shape tends towards $|\sin\omega t|$ in even quarters of
each period and zero in odd ones. Such a signal has
$P_2/P_{\mathrm{source}}=45.0\%$, which is indeed the asymptotic
efficiency at $R_{off}/R_{on}\rightarrow \infty$. We recognize
this signal shape as being \emph{two-quarter-wave rectified}. For
higher harmonics generation, we find the value after optimization
of $(\sum_{k=2}^\infty P_k)/P_{\mathrm{source}}$ to be 56.4\%, a
modest improvement over the single device circuit.

In conclusion, we have demonstrated the potential of memristive
systems for passive second and higher harmonics generation. Using
an approach to maximize the rate of power conversion for a
specific harmonic, we have shown that memristive circuits are much
more efficient for harmonic generation purposes (at optimal
operation conditions) than the traditional diode bridge. In
addition, the operation voltage in memristive circuits can be
lower than that used in diode circuits because of $\approx$ 0.7V
barrier voltage of silicon p-n junctions~\cite{Alexander08a}. We
also anticipate that memristive devices can be beneficially used
in harmonics generation in combination with active circuits. An
example of such a circuit is a memristive bridge operating with a
high resistance load followed by an operational amplifier.
Finally, memcapacitive and meminductive
systems~\cite{diventra09a,pershin11a} can be used for passive (and
low-dissipative) higher harmonics generation instead of
memristive ones. The results of our investigation can be readily
tested experimentally.

This work has been partially supported by NSF grant No.
DMR-0802830 and the Center for Magnetic Recording Research at
UCSD.

% Write about the $0.7V->1.4V$ problem in
% the diode bridge that does not exist here. Write the output is
% $V_0|\sin\omega t|\theta(V_0|\sin\omega t|-1.4)$. summary and
%suggestions for future research and experimental verification.
%Consider more citations. Change $R_{\mathrm{on}}$ to $R_{\mathrm{on}}$.
%Check grammar and wording.\\

%


\begin{thebibliography}{10}%
\makeatletter
\providecommand \@ifxundefined [1]{%
 \ifx #1\undefined \expandafter \@firstoftwo
 \else \expandafter \@secondoftwo
\fi
}%
\providecommand \@ifnum [1]{%
 \ifnum #1\expandafter \@firstoftwo
 \else \expandafter \@secondoftwo
\fi
}%
\providecommand \enquote [1]{``#1''}%
\providecommand \bibnamefont  [1]{#1}%
\providecommand \bibfnamefont [1]{#1}%
\providecommand \citenamefont [1]{#1}%
\providecommand\href[0]{\@sanitize\@href}%
\providecommand\@href[1]{\endgroup\@@startlink{#1}\endgroup\@@href}%
\providecommand\@@href[1]{#1\@@endlink}%
\providecommand \@sanitize [0]{\begingroup\catcode`\&12\catcode`\#12\relax}%
\@ifxundefined \pdfoutput {\@firstoftwo}{%
 \@ifnum{\z@=\pdfoutput}{\@firstoftwo}{\@secondoftwo}%
}{%
 \providecommand\@@startlink[1]{\leavevmode}%
 \providecommand\@@endlink[0]{}%
}{%
 \providecommand\@@startlink[1]{%
  \leavevmode
  \pdfstartlink
   attr{/Border[0 0 1 ]/H/I/C[0 1 1]}%
   user{/Subtype/Link/A<</Type/Action/S/URI/URI(#1)>>}%
  \relax
 }%
 \providecommand\@@endlink[0]{\pdfendlink}%
}%
\providecommand \url  [0]{\begingroup\@sanitize \@url }%
\providecommand \@url [1]{\endgroup\@href {#1}{\urlprefix}}%
\providecommand \urlprefix [0]{URL }%
\providecommand \Eprint[0]{\href }%
\@ifxundefined \urlstyle {%
  \providecommand \doi [1]{doi:\discretionary{}{}{}#1}%
}{%
  \providecommand \doi [0]{doi:\discretionary{}{}{}\begingroup
  \urlstyle{rm}\Url }%
}%
\providecommand \doibase [0]{http://dx.doi.org/}%
\providecommand \Doi[1]{\href{\doibase#1}}%
\providecommand \bibAnnote [3]{%
  \BibitemShut{#1}%
  \begin{quotation}\noindent
    \textsc{Key:}\ #2\\\textsc{Annotation:}\ #3%
  \end{quotation}%
}%
\providecommand \bibAnnoteFile [2]{%
  \IfFileExists{#2}{\bibAnnote {#1} {#2} {\input{#2}}}{}%
}%
\providecommand \typeout [0]{\immediate \write \m@ne }%
\providecommand \selectlanguage [0]{\@gobble}%
\providecommand \bibinfo [0]{\@secondoftwo}%
\providecommand \bibfield [0]{\@secondoftwo}%
\providecommand \translation [1]{[#1]}%
\providecommand \BibitemOpen[0]{}%
\providecommand \bibitemStop [0]{}%
\providecommand \bibitemNoStop [0]{.\EOS\space}%
\providecommand \EOS [0]{\spacefactor3000\relax}%
\providecommand \BibitemShut [1]{\csname bibitem#1\endcsname}%
%</preamble>
\bibitem{Franken61a}%
  \BibitemOpen
  \bibfield{author}{%
  \bibinfo {author} {\bibfnamefont{P.~A.}\ \bibnamefont{Franken}}, \bibinfo
  {author} {\bibfnamefont{A.~E.}\ \bibnamefont{Hill}}, \bibinfo {author}
  {\bibfnamefont{C.~W.}\ \bibnamefont{Peters}},\ and\ \bibinfo {author}
  {\bibfnamefont{G.}~\bibnamefont{Weinreich}},\ }%
  \bibfield{journal}{%
  \bibinfo {journal} {Phys. Rev. Lett.}\ }%
  \textbf{\bibinfo {volume} {7}},\ \bibinfo {pages} {118} (\bibinfo {year}
  {1961})%
  \bibAnnoteFile{NoStop}{Franken61a}%
\bibitem{Hecht01a}%
  \BibitemOpen
  \bibfield{author}{%
  \bibinfo {author} {\bibfnamefont{E.}~\bibnamefont{Hecht}},\ }%
  \emph{\bibinfo {title} {Optics}},\ \bibinfo {edition} {4th}\ ed.\ (\bibinfo
  {publisher} {Addison Wesley},\ \bibinfo {year} {2001})%
  \bibAnnoteFile{NoStop}{Hecht01a}%
\bibitem{Ou92a}%
  \BibitemOpen
  \bibfield{author}{%
  \bibinfo {author} {\bibfnamefont{Z.}~\bibnamefont{Ou}}, \bibinfo {author}
  {\bibfnamefont{S.}~\bibnamefont{Pereira}}, \bibinfo {author}
  {\bibfnamefont{E.}~\bibnamefont{Polzik}},\ and\ \bibinfo {author}
  {\bibfnamefont{H.}~\bibnamefont{Kimble}},\ }%
  \bibfield{journal}{%
  \bibinfo {journal} {Opt. Lett.}\ }%
  \textbf{\bibinfo {volume} {17}},\ \bibinfo {pages} {640} (\bibinfo {year}
  {1992})%
  \bibAnnoteFile{NoStop}{Ou92a}%
\bibitem{pershin09a}%
  \BibitemOpen
  \bibfield{author}{%
  \bibinfo {author} {\bibfnamefont{Y.~V.}\ \bibnamefont{Pershin}}\ and\
  \bibinfo {author} {\bibfnamefont{M.}~\bibnamefont{{Di Ventra}}},\ }%
  \bibfield{journal}{%
  \bibinfo {journal} {Phys. Rev. B}\ }%
  \textbf{\bibinfo {volume} {79}},\ \bibinfo {pages} {153307} (\bibinfo {year}
  {2009})%
  \bibAnnoteFile{NoStop}{pershin09a}%
\bibitem{Alexander08a}%
  \BibitemOpen
  \bibfield{author}{%
  \bibinfo {author} {\bibfnamefont{C.}~\bibnamefont{Alexander}}\ and\ \bibinfo
  {author} {\bibfnamefont{M.}~\bibnamefont{Sadiku}},\ }%
  \emph{\bibinfo {title} {Fundamentals of Electric Circuits}},\ \bibinfo
  {edition} {4th}\ ed.\ (\bibinfo {publisher} {McGraw-Hill},\ \bibinfo {year}
  {2008})%
  \bibAnnoteFile{NoStop}{Alexander08a}%
\bibitem{Chua71a}%
  \BibitemOpen
  \bibfield{author}{%
  \bibinfo {author} {\bibfnamefont{L.~O.}\ \bibnamefont{Chua}},\ }%
  \bibfield{journal}{%
  \bibinfo {journal} {{IEEE} Trans. Circuit Theory}\ }%
  \textbf{\bibinfo {volume} {18}},\ \bibinfo {pages} {507} (\bibinfo {year}
  {1971})%
  \bibAnnoteFile{NoStop}{Chua71a}%
\bibitem{chua76a}%
  \BibitemOpen
  \bibfield{author}{%
  \bibinfo {author} {\bibfnamefont{L.~O.}\ \bibnamefont{Chua}}\ and\ \bibinfo
  {author} {\bibfnamefont{S.~M.}\ \bibnamefont{Kang}},\ }%
  \bibfield{journal}{%
  \bibinfo {journal} {Proc. {IEEE}}\ }%
  \textbf{\bibinfo {volume} {64}},\ \bibinfo {pages} {209} (\bibinfo {year}
  {1976})%
  \bibAnnoteFile{NoStop}{chua76a}%
\bibitem{Strukov08a}%
  \BibitemOpen
  \bibfield{author}{%
  \bibinfo {author} {\bibfnamefont{D.~B.}\ \bibnamefont{Strukov}}, \bibinfo
  {author} {\bibfnamefont{G.~S.}\ \bibnamefont{Snider}}, \bibinfo {author}
  {\bibfnamefont{D.~R.}\ \bibnamefont{Stewart}},\ and\ \bibinfo {author}
  {\bibfnamefont{R.~S.}\ \bibnamefont{Williams}},\ }%
  \bibfield{journal}{%
  \bibinfo {journal} {Nature}\ }%
  \textbf{\bibinfo {volume} {453}},\ \bibinfo {pages} {80} (\bibinfo {year}
  {2008})%
  \bibAnnoteFile{NoStop}{Strukov08a}%
\bibitem{pershin11a}%
  \BibitemOpen
  \bibfield{author}{%
  \bibinfo {author} {\bibfnamefont{Y.~V.}\ \bibnamefont{Pershin}}\ and\
  \bibinfo {author} {\bibfnamefont{M.}~\bibnamefont{Di~Ventra}},\ }%
  \bibfield{journal}{%
  \bibinfo {journal} {Adv. Phys.}\ }%
  \textbf{\bibinfo {volume} {60}},\ \bibinfo {pages} {145} (\bibinfo {year}
  {2011})%
  \bibAnnoteFile{NoStop}{pershin11a}%
\bibitem{diventra09a}%
  \BibitemOpen
  \bibfield{author}{%
  \bibinfo {author} {\bibfnamefont{M.}~\bibnamefont{{Di Ventra}}}, \bibinfo
  {author} {\bibfnamefont{Y.~V.}\ \bibnamefont{Pershin}},\ and\ \bibinfo
  {author} {\bibfnamefont{L.~O.}\ \bibnamefont{Chua}},\ }%
  \bibfield{journal}{%
  \Doi{10.1109/JPROC.2009.2021077}{\bibinfo {journal} {Proc. {IEEE}}}\ }%
  \textbf{\bibinfo {volume} {97}},\ \bibinfo {pages} {1717} (\bibinfo {year}
  {2009})%
  \bibAnnoteFile{NoStop}{diventra09a}%
\bibitem{Oskoee11}%
  \BibitemOpen
  \bibfield{author}{%
  \bibinfo {author} {\bibfnamefont{E.}~\bibnamefont{Oskoee}}\ and\ \bibinfo
  {author} {\bibfnamefont{M.}~\bibnamefont{Sahimi}},\ }%
  \bibfield{journal}{%
  \bibinfo {journal} {Phys. Rev. E}\ }%
  \textbf{\bibinfo {volume} {83}},\ \bibinfo {pages} {031105} (\bibinfo {year}
  {2011})%
  \bibAnnoteFile{NoStop}{Oskoee11}%
\bibitem{yang08a}%
  \BibitemOpen
  \bibfield{author}{%
  \bibinfo {author} {\bibfnamefont{J.~J.}\ \bibnamefont{Yang}}, \bibinfo
  {author} {\bibfnamefont{M.~D.}\ \bibnamefont{Pickett}}, \bibinfo {author}
  {\bibfnamefont{X.}~\bibnamefont{Li}}, \bibinfo {author}
  {\bibfnamefont{D.~A.~A.}\ \bibnamefont{Ohlberg}}, \bibinfo {author}
  {\bibfnamefont{D.~R.}\ \bibnamefont{Stewart}},\ and\ \bibinfo {author}
  {\bibfnamefont{R.~S.}\ \bibnamefont{Williams}},\ }%
  \bibfield{journal}{%
  \bibinfo {journal} {Nat. Nanotechnol.}\ }%
  \textbf{\bibinfo {volume} {3}},\ \bibinfo {pages} {429} (\bibinfo {year}
  {2008})%
  \bibAnnoteFile{NoStop}{yang08a}%
\bibitem{Baikalov03a}%
  \BibitemOpen
  \bibfield{author}{%
  \bibinfo {author} {\bibfnamefont{A.}~\bibnamefont{Baikalov}}, \bibinfo
  {author} {\bibfnamefont{Y.}~\bibnamefont{Wang}}, \bibinfo {author}
  {\bibfnamefont{B.}~\bibnamefont{Shen}}, \bibinfo {author}
  {\bibfnamefont{B.}~\bibnamefont{Lorenz}}, \bibinfo {author}
  {\bibfnamefont{S.}~\bibnamefont{Tsui}}, \bibinfo {author}
  {\bibfnamefont{Y.}~\bibnamefont{Sun}}, \bibinfo {author}
  {\bibfnamefont{Y.}~\bibnamefont{Xue}},\ and\ \bibinfo {author}
  {\bibfnamefont{C.}~\bibnamefont{Chu}},\ }%
  \bibfield{journal}{%
  \bibinfo {journal} {Appl. Phys. Lett.}\ }%
  \textbf{\bibinfo {volume} {83}},\ \bibinfo {pages} {957} (\bibinfo {year}
  {2003})%
  \bibAnnoteFile{NoStop}{Baikalov03a}%
\bibitem{Tsui04a}%
  \BibitemOpen
  \bibfield{author}{%
  \bibinfo {author} {\bibfnamefont{S.}~\bibnamefont{Tsui}}, \bibinfo {author}
  {\bibfnamefont{A.}~\bibnamefont{Baikalov}}, \bibinfo {author}
  {\bibfnamefont{J.}~\bibnamefont{Cmaidalka}}, \bibinfo {author}
  {\bibfnamefont{Y.}~\bibnamefont{Sun}}, \bibinfo {author}
  {\bibfnamefont{Y.}~\bibnamefont{Wang}}, \bibinfo {author}
  {\bibfnamefont{Y.}~\bibnamefont{Yue}}, \bibinfo {author}
  {\bibfnamefont{C.}~\bibnamefont{Chu}}, \bibinfo {author}
  {\bibfnamefont{L.}~\bibnamefont{Chen}},\ and\ \bibinfo {author}
  {\bibfnamefont{A.}~\bibnamefont{Jacobson}},\ }%
  \bibfield{journal}{%
  \bibinfo {journal} {Appl. Phys. Lett.}\ }%
  \textbf{\bibinfo {volume} {85}} (\bibinfo {year} {2004})%
  \bibAnnoteFile{NoStop}{Tsui04a}%
\bibitem{Munson10a}%
  \BibitemOpen
  \bibfield{author}{%
  \bibinfo {author} {\bibfnamefont{B.~R.}\ \bibnamefont{Munson}}, \bibinfo
  {author} {\bibfnamefont{D.}~\bibnamefont{Young}}, \bibinfo {author}
  {\bibfnamefont{T.}~\bibnamefont{Okiishi}},\ and\ \bibinfo {author}
  {\bibfnamefont{W.}~\bibnamefont{Huebsch}},\ }%
  \emph{\bibinfo {title} {Fundamentals of Fluid Mechanics}},\ \bibinfo
  {edition} {6th}\ ed.\ (\bibinfo {publisher} {Wiley},\ \bibinfo {year}
  {2010})%
  \bibAnnoteFile{NoStop}{Munson10a}%
\bibitem{Wang10b}%
  \BibitemOpen
  \bibfield{author}{%
  \bibinfo {author} {\bibfnamefont{F.~Z.}\ \bibnamefont{Wang}}, \bibinfo
  {author} {\bibfnamefont{N.}~\bibnamefont{Helian}}, \bibinfo {author}
  {\bibfnamefont{S.}~\bibnamefont{Wu}}, \bibinfo {author}
  {\bibfnamefont{M.-G.}\ \bibnamefont{Lim}}, \bibinfo {author}
  {\bibfnamefont{Y.}~\bibnamefont{Guo}},\ and\ \bibinfo {author}
  {\bibfnamefont{M.~A.}\ \bibnamefont{Parker}},\ }%
  \bibfield{journal}{%
  \bibinfo {journal} {El. Dev. Lett.}\ }%
  \textbf{\bibinfo {volume} {31}},\ \bibinfo {pages} {755} (\bibinfo {year}
  {2010})%
  \bibAnnoteFile{NoStop}{Wang10b}%
\end{thebibliography}
\end{document}